\documentclass[pdflatex,sn-mathphys-num]{sn-jnl}

\usepackage{graphicx}%
\usepackage{multirow}%
\usepackage{amsmath,amssymb,amsfonts}%
\usepackage{amsthm}%
\usepackage{mathrsfs}%
\usepackage[title]{appendix}%
\usepackage{xcolor}%
\usepackage{textcomp}%
\usepackage{manyfoot}%
\usepackage{booktabs}%
\usepackage{algorithm}%
\usepackage{algorithmicx}%
\usepackage{algpseudocode}%
\usepackage{listings}%

\usepackage{graphicx, bm}

\usepackage{graphicx}
\usepackage{tikz}
\usetikzlibrary{arrows, shapes, positioning, calc}
\usepackage{quantikz}

\theoremstyle{thmstyleone}%
\theoremstyle{thmstyletwo}%

\theoremstyle{thmstylethree}%

\begin{document}


\title{ Enhanced Variational Quantum Kolmogorov-Arnold Network }

\author*[1]{\fnm{ Hikaru } \sur{ Wakaura }}\email{ hikaruwakaura@gmail.com }

\author[2, 3]{ Rahmat Mulyawan }  \email{ rahmat.mulyawan@itb.ac.id } 
 
\author[ 2, 3, 4] {\fnm{ Andriyan  } \sur{ B. Suksmono }}\email{  suksmono@itb.ac.id }
    
 \affil[1]{ QuantScape Inc. QuantScape Inc., 4-11-18, Manshon-Shimizudai, Meguro, Tokyo, 153-0064, Japan }  
             
\affil[2]{ The School of Electrical Engineering and Informatics, Institut Teknologi Bandung (STEI-ITB), Jl. Ganesha No.10, Bandung, Indonesia }  
      
\affil[3]{ Research Collaboration Center for Quantum Technology 2.0, BRIN-ITB-TelU, Indonesia }    
       
\affil[4]{ ITB Research Center on ICT (PPTIK-ITB) }

  
\abstract{ The Kolmogorov-Arnold Network (KAN) is a multi-layer network model in which the synapses between neurons, rather than the neurons themselves, carry the trainable functions. Existing quantum implementations of KAN either lack accuracy (Variational Quantum KAN, VQKAN) or rely on block encoding and Quantum Signal Processing, which demand large numbers of control gates and ancillae. We propose the Enhanced Variational Quantum Kolmogorov-Arnold Network (EVQKAN), a variational ansatz that emulates a $ 2^{N_q} $-dimensional KAN layer matrix by tiling controlled rotations through a sum-operator construction, and that requires only $ 2^{N_q-1} $ trainable spline functions per layer. On the task of fitting an elementary function, EVQKAN attains a lower test error than Quantum Neural Networks (QNN), VQKAN and Adaptive VQKAN, and the separation is statistically significant in every case ( Mann-Whitney $ p < 0.002 $, Cliff's $ \delta \leq -0.86 $ over ten independent attempts; EVQKAN beats VQKAN on every attempt ), though classical KAN remains more accurate than every quantum method we tested. On a two-dimensional classification task the picture reverses: under a leak-free protocol that we introduce here, EVQKAN classifies above chance ( accuracy $ 0.620 $, $ p = 0.0005 $ ) but is significantly \emph{less} accurate than a QNN carrying one fifth as many parameters ( $ \Delta $ accuracy $ -0.134 $, $ p = 0.0014 $; $ \Delta $ AUC $ -0.252 $, $ p = 0.0002 $ ). The classification results of an earlier version of this work are withdrawn: their input encoding placed the target label into the circuit as a feature for EVQKAN but not for the methods it was compared against. We further show that the dominant error source is overfitting driven by an under-determined training set: sweeping the number of training samples closes the train-test gap by $ 58\% $ ( Spearman $ p < 10^{-3} $ ). We report the circuit cost honestly --- three layers emit $ 1017 $ operations, or $ 4110 $ two-qubit gates once the multi-controlled gates are decomposed --- so the present construction is a simulator-scale and fault-tolerant-era proposal rather than a NISQ-ready one, and we identify block encoding and qubitization as the route to reducing it. }


  
 
\maketitle

\section{Introduction}\label{1}

The rapid advancements in artificial intelligence (AI) have been largely driven by neural network models inspired by the structure of the human brain \cite{Yu2019}. These models, composed of interconnected artificial neurons or perceptrons \cite{https://doi.org/10.1049/cit2.12286, https://doi.org/10.1049/cit2.12014}, have demonstrated remarkable success in a wide range of applications, including image recognition and natural language processing \cite{https://doi.org/10.1049/trit.2018.1008}. However, conventional neural networks face significant challenges in terms of scalability and computational efficiency when processing large-scale data, which limits the continued advancement of AI technologies.

To address these issues, alternative network architectures have been explored. One particularly promising approach is the Kolmogorov–Arnold Network (KAN), recently proposed by Tegmark's group \cite{2024arXiv240419756L}. 

KAN enhances computational efficiency by directly optimizing the functions of synaptic weights, instead of neuron parameters, using matrix operations for streamlined computation. 

Moreover, its architecture can be naturally interpreted and implemented as a quantum circuit, paving the way for integration with quantum computing.
  
As a result, KAN has attracted growing global interest, with numerous studies exploring its theoretical foundations and practical applications. Despite some critical perspectives \cite{2024arXiv241106727C}, a broad range of research has been reported on KAN, including its application to image analysis \cite{2024arXiv241118165H}, time-dependent data \cite{2024arXiv241203710K}, and physical modeling tasks \cite{2024arXiv241008452B, 2024arXiv241114902K}. KAN has also been applied in domains such as spacecraft control and medical diagnostics \cite{2024arXiv241007446J, 2024arXiv240800273T}.      
     
Since its introduction, efforts have been made to implement KAN on quantum computers. Quantum computers, first proposed by Richard P. Feynman \cite{feynman_simulating_1982}, exploit quantum superposition and entanglement to solve certain problems exponentially faster than classical computers. The computational power of a quantum system comes from its ability to represent data in a \( 2^{N_q} \)-dimensional Hilbert space, where \( N_q \) is the number of qubits. This makes KAN a strong candidate for efficient implementation on quantum hardware, particularly for large multi-layer networks.
         
Several quantum implementations of KAN have been proposed. Variational Quantum KAN (VQKAN) \cite{Wakaura_VQKAN_2024} adapts KAN to a variational quantum framework, using qubit measurements as neuron activations and quantum gates as synaptic functions. Another approach, Quantum KAN \cite{2024arXiv241004435I}, leverages Quantum Signal Processing \cite{2024PRXQ....5b0368M} and block encoding techniques \cite{Sunderhauf2024blockencoding} to represent KAN layers in quantum circuits.

Additionally, methods like Quantum Architecture Search \cite{2024EPJQT..11...76K} have been proposed to optimize the quantum circuit structure of KAN. However, current implementations face several limitations. VQKAN suffers from insufficient accuracy for practical use, while Quantum KAN requires a large number of control gates and ancillary qubits, making it unsuitable for today’s noisy intermediate-scale quantum (NISQ) devices. Moreover, the optimization of Quantum KAN becomes increasingly demanding with network size, due to the need to train a number of parameters proportional to the number of elements in each layer matrix.
 
To address these challenges, we propose a novel quantum extension of KAN called Enhanced VQKAN (EVQKAN), designed within the framework of Variational Quantum Algorithms (VQAs). VQKAN reformulates the KAN structure using parametric quantum gates and feedback loops, emulating the original architecture within a VQE-style scheme. EVQKAN improves upon VQKAN by employing matrix-based layers that mimic KAN behavior, requiring only \( 2^{N_q - 1} \) trainable spline functions for a \( 2^{N_q} \)-dimensional layer. This is the sense in which EVQKAN is parameter-efficient: the number of trainable functions grows as the square root of the layer dimension, whereas Quantum KAN trains a number of parameters proportional to the number of elements of the layer matrix.

We stress at the outset that this parameter economy does \emph{not} translate into a small circuit. The tiled sum-operator construction is realised with multi-controlled rotations, and for \( N_q = 3 \) a three-layer EVQKAN emits \( 1017 \) operations, of which \( 684 \) are Toffoli gates; decomposing each Toffoli into six CNOTs and nine one-qubit gates gives \( 4110 \) two-qubit gates and \( 10593 \) gates in total. Circuits of this depth are beyond the reach of present-day hardware. EVQKAN is therefore best understood as a simulator-scale and fault-tolerant-era construction, and all results reported below are noiseless state-vector simulations. Reducing the gate count --- for which block encoding and qubitization are the natural tools --- is the principal obstacle between the present work and hardware deployment, and we return to it in Section \ref{4}.

Quantum computing algorithms, particularly VQAs, have advanced rapidly in recent years. Foundational work by Aspuru-Guzik and collaborators \cite{Kassal2011} has led to the development of widely used algorithms such as the Variational Quantum Eigensolver (VQE) \cite{McClean_2016}, Adaptive VQE \cite{2019NatCo..10.3007G}, and Multiscale Contracted VQE (MCVQE) \cite{2019arXiv190608728P}, among others \cite{2021arXiv210501141W, 2021arXiv210902009W}. These algorithms are well-suited for NISQ devices and have been successfully applied to quantum machine learning tasks \cite{2014PhRvL.113m0503R, 2019QS&T....4a4001K, 2019Natur.567..209H, 2022PhRvA.106b2601A, 2022arXiv220211200K, 2021PhRvP..16d4057B, 2020PhRvL.125j0401W, 2022arXiv220608316Y, PhysRevA.98.032309}.
    
We evaluate EVQKAN on two tasks. On the fitting of elementary functions EVQKAN attains a significantly lower test error than QNN \cite{mcclean_barren_2018}, VQKAN and Adaptive VQKAN, with large effect sizes over ten independent attempts, while classical KAN remains more accurate than every quantum method we tested. On a two-dimensional classification task the ordering reverses and EVQKAN falls significantly behind QNN. We report both outcomes, because the contrast between them is itself the most informative result: the advantage the tiled ansatz confers on function fitting does not transfer, and we have not been able to attribute the failure to the readout, the parameter count or the optimiser budget.

The contribution of this paper is therefore an architecture that makes variational quantum KAN competitive \emph{among quantum models} on function fitting, a leak-free protocol and a set of standard metrics under which quantum KAN variants can be compared honestly, and a quantitative account of what currently limits the method --- overfitting from an under-determined training set, and a circuit cost dominated by multi-controlled gates.

Section \ref{1} is the introduction, section \ref{2} describes the method detail of EVQKAN and the optimization method, section \ref{3} describes the results of the fitting and classification problems, Section \ref{4} is the discussion of results, and section \ref{7} is the concluding remark.

\section{Method}\label{2}

 In this section, we describe the method and implementation of the Enhanced Variational Quantum Kolmogorov-Arnold Network (EVQKAN).
VQKAN is the variational quantum algorithm version of KAN, a multi-layer network based on the connection of synapses in neurons. 
 First, initial state $ \mid \Psi_{ini} (_1 {\bf x} ^m) \rangle $ is $ \prod _{j = 0} ^{N _q -1 } Ry^j (acos(2 _1 {\bf x}_j ^m -1)) \mid 0 \rangle ^{\otimes N _q} $ for each input $ m $. $ Ry^{j} (\theta) $ is $ \theta $ degrees angle rotation gate for y-axis on qubit $ j $. 
$ _n {\bf x} ^m $ is the input vector at layer n for $ m $-th input data.  
 We will describe later that the Loss function is calculated similarly to Subspace-search VQE \cite{Nakanishi2018a}, and multiple points are calculated at once. 
 The rotation angles fed into the ansatz are generated by trainable spline functions, as in classical KAN. This is the point at which EVQKAN departs from VQKAN, and it is the source of its parameter economy, so we state it precisely.

Layer $ n $ carries $ N_f = 2^{N_q-1} $ trainable spline functions $ g_j^n $, $ j = 0, \ldots, N_f-1 $, each defined on $ [0,1] $ and represented by $ N_g $ control points $ c_s^{n j} $,
\begin{equation}
g_j^{n} ( u ) = \sum_{s = 0}^{N_g - 1} c_s^{n j} B_s ( u ),
\end{equation}
where the $ B_s $ are the B-spline basis functions and the $ c_s^{n j} $ are initialised to zero. The layer receives $ N_d^n = \dim ( _n {\bf x} ^m ) $ readout values $ _n x_w ^m \in [0,1] $, and the angle applied in tile $ j $ at position $ w $ is obtained by evaluating the $ j $-th spline at the $ w $-th readout,
\begin{equation}
\phi_{j w}^{n} ( _n {\bf x} ^m ) = 2 \arccos \! \left( \frac{ g_j^n ( _n x_w ^m ) + E_f ( _n x_w ^m ) - g_{j,\min}^n }{ g_{j,\max}^n + E_f ( 1 ) - g_{j,\min}^n } \right),
\label{ phieq }
\end{equation}
with $ g_{j,\min}^n $ and $ g_{j,\max}^n $ the extrema of $ g_j^n $ over $ [0,1] $, which keep the argument of $ \arccos $ inside its domain.

Two consequences are worth making explicit, because they are easy to misread. First, the angle carries two indices $ ( j, w ) $ and a layer therefore consumes $ N_f \times N_d^n $ angles --- sixteen for the $ N_q = 3, N_d^n = 4 $ configuration used throughout this paper --- but the \emph{trainable} coefficients carry only the single index $ j $. A layer holds $ N_f N_g = 2^{N_q-1} N_g $ free parameters, not $ N_f N_d^n N_g $. Second, only one element of $ _n {\bf x} ^m $ enters each $ \phi_{j w}^{n} $, in contrast to VQKAN where the angle sums over all elements of the input vector.

For $ N_q = 3 $, $ N_g = 8 $ and three layers this gives $ 4 \times 8 \times 3 = 96 $ trainable parameters in total. We note that an earlier version of this work optimised a $ 384 $-component vector for the same model, of which only these $ 96 $ components were ever read; the redundant coordinates left the objective exactly invariant while consuming the optimiser's budget. All results below use the $ 96 $-dimensional parameterisation.

$ E_f (_n x_i ^m) = _n x_i ^m / (exp (-_n x_i ^m) + 1) $ is the Fermi-Dirac expectation energy-like value of the distribution. The component of $ _n {\bf x} $ is the expectation value of the given observable for the calculated states of qubits.  
The layer corresponds to the quantum circuit to make a superposition state called ansatz.  
EVQKAN uses the tiled matrices by the elements of gate operation matrices made by the sum operator and block encoding technics. 
We use the method of sum operator for tiling, like,
  
 \begin{equation}   
 U ^{ k, \{ 0, 2 ^ { k }-1 \} } _n = U ^{ k-1, \{ 0, 2 ^ { k-1 }-1 \} } _n + X_k U ^{ k-1, \{ 2 ^ { k-1 }, 2 ^ { k }-1 \} } _n 
 \end{equation} 
\begin{equation}   
  U ^ { 0, p }_n = \prod_{ j = 0 }^{ 2^ { N_q - 1 } -1 } C ^ { N_q-1 }_{ j } Ry^0( \phi_{ j p }^n ( _n {\bf x} ^m _{ j } ) )      
\end{equation}                     
        
Then, $ C ^ { k }_{ j } Ry^0 $ is the k-qubit controlled y-axis rotation gate with 0 th qubit as target qubit that acts y-axis rotation gate when the control qubits are $ \mid j \rangle $ state for the decimal expression of binary state of qubits and $ N_q $ is the number of qubits, respectively.                    
                    
The entire ansatz is,          
            
\begin{equation}     
{\Phi}_n^E = M U ^{ N_q-1, \{ 0, 2 ^ { N_q-1 }-1 \} } _n \label{ Phia }
 \end{equation}        
    
\begin{equation}      
\mid \Psi ( _1 {\bf x} ^m ) \rangle = \prod_{ n = 1 }^{ num. ~ of ~ layers ~ N_l} { \Phi}_n^{ E }M \mid \Psi_{ ini } ( _1 {\bf x} ^m ) \rangle    
\end{equation}     
     
We implement the technique for implementation of sum operators by the manners on paper \cite{PhysRevResearch.2.033043}. 
Then, $ M $ indicates the measurement of all qubits and deriving the $ _{ n + 1 } {\bf x} ^m $s. 
  
\begin{figure}     
\centering  
 
\includegraphics[scale= 0.3 ]{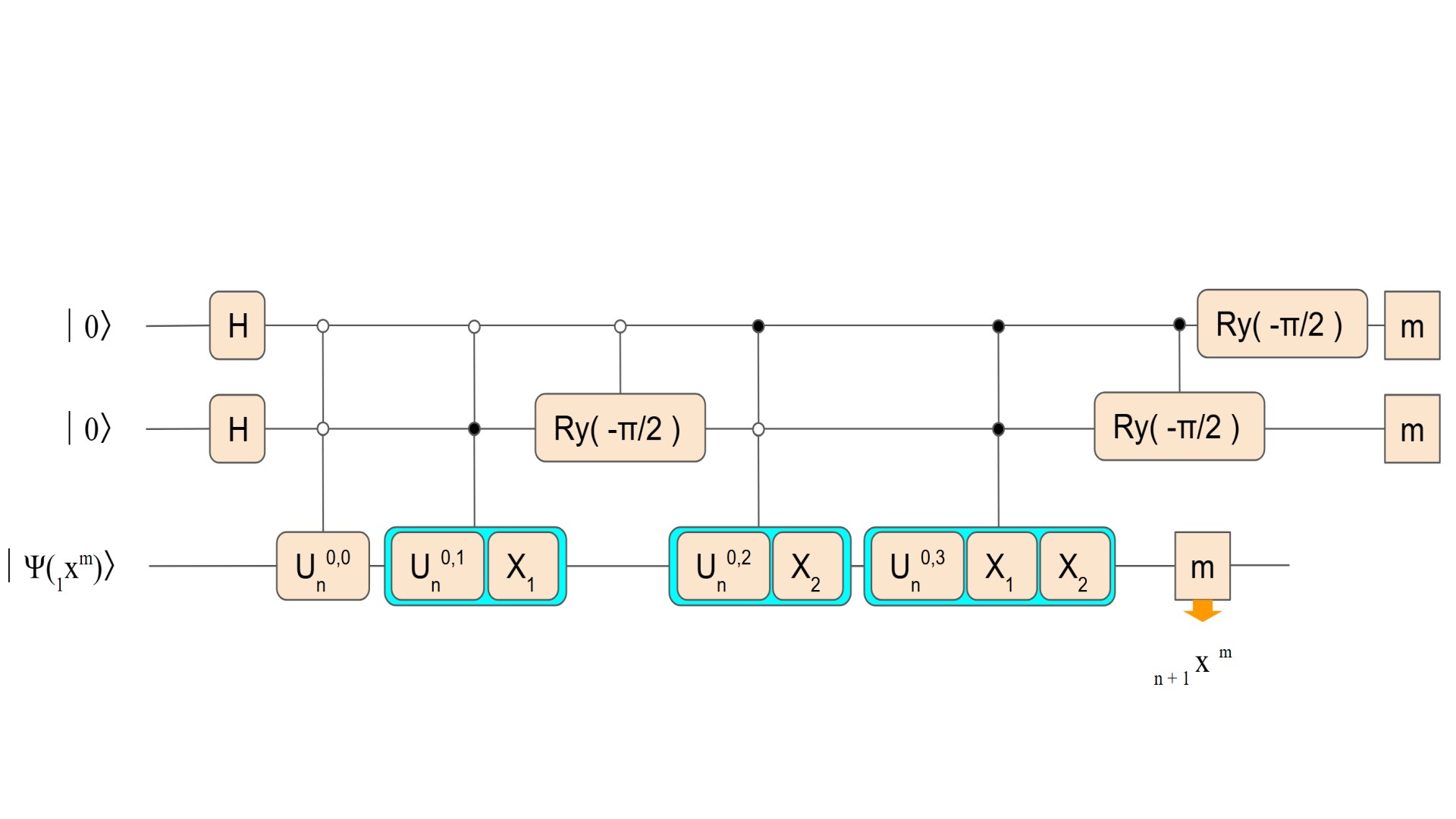} 
 
\caption{   
Simplified picture of our ansatz on EVQKAN.  
White circle indicates that connected operators are acted on the circuit in case the qubits that the circle exists is $ \mid 0 \rangle $ state and black circle indicates that connected operators are acted on the circuit in case the qubits that the circle exists is $ \mid 1 \rangle $ state, respectively.  
} \label{ e v q }   
\end{figure}

The illustrated circuit of a single layer is as Fig.\ref{ e v q }.    
We will demonstrate solving fitting problems of elementary equations using the example of $ N_q = 3 $.      
Sum operators are gained in case the measurement result of ancillary qubits are all zero state.          
For working qubits, the state of the qubits is destroyed after measurement, and calculations commence from scratch using the measured input vector, omitting $ m $ gates in case calculations are done on real devices.  
We save the states for easiness because the demonstrations on the paper are all numerical simulations.       
  
Others are the same as ordinary VQKAN \cite{Wakaura_VQKAN_2024}.

 The result is readout as a form of the Hamiltonian expectation value $ H $, and the loss function is calculated as follows,
\begin{eqnarray} 
l_m &=& | \langle \Psi ( _1 {\bf x} ^m ) | H | \Psi ( _1 {\bf x} ^m ) \rangle-f^{ aim } ( _1 {\bf x} ^m ) | \\\label{ loss } 
 L &=& \sum_{ m = 0 }^{ num. ~ of ~ samples ~ N -1 } a_m l_m \\\nonumber
\end{eqnarray} 
where $ f^{ aim } ( _1 {\bf x} ^m ) $ is the aimed value of sampled point m and $ l_m $ is the loss function ( absolute distance ) of point $  m $, respectively. Hamiltonian takes the form $H = \sum_{j = 0}^{N_o-1 }  \theta_j P_j $ for the product of the Pauli matrix $P_j$, consisting of the Pauli matrix $X_j, Y_j, Z_j$. $ N_o $ is the number of $P_j$ in Hamiltonian. 

The condition of convergence is the default of scipy for all methods. We assume $ N = 10, a_m = ( N-m ) /N, N_l = 3, N_q = 3 $ and $ N_g = 8 $, respectively.

The spline defined by the $ N_g $ control points is sampled on a fixed grid of $ N_s $ points before the value at $ _n x_w ^m $ is read off by nearest-neighbour lookup. We hold $ N_s $ fixed for the whole optimisation. This is worth stating because an earlier version of this work incremented $ N_s $ on every optimiser callback, which changes the objective function between iterations and invalidates the trust-region model that COBYLA builds; the results reported here do not do this, and all arms of every comparison are optimised against the same fixed function.

We use blueqat SDK \cite{Kato} for numerical simulation of quantum calculations and COBYLA of scipy to optimize parameters unless the use of another optimizer is declared.

All results in this paper are noiseless state-vector simulations, and we assume that the number of shots is infinite. No noise model of any kind is applied. Consequently nothing reported here constrains the behaviour of EVQKAN under decoherence or finite sampling; see the limitations discussed in Section \ref{4}.

All calculations except where declared are performed in Jupyter notebook with Anaconda 3.9.12 and Intel Core i7-9750H.
   
\section{Result} \label{3}   
   
In this section, we show the EVQKAN result on fitting the elemental equation.
Each layer consumes 16 angles $ \phi_{j w} $ generated from 4 trainable spline functions, giving 96 trainable parameters for the three-layer network, and the number of qubits, including ancillae, is 8.
We performed the EVQKAN of the following equation on 10 sampled points and predicted the values of 50 test points after optimization by sampled points.

Throughout, every method is run for ten independent attempts. We report the mean, the sample standard deviation and the 95\% confidence interval of the mean, and we test differences with the two-sided Mann-Whitney $ U $ test, quoting Cliff's $ \delta $ as the effect size. The attempts are independent runs rather than paired seeds, which is why an unpaired test is appropriate. We adopt this convention because the ranges of the methods compared here overlap substantially, so a comparison of means alone cannot distinguish a real separation from run-to-run scatter.

 \subsection{ Fitting problem }   
First, we describe the result of the fitting problem.  
The target function is defined as:   
  
\begin{equation}   
 f^{\rm aim} ({\bf x}) = \exp\left(\sin(x_0^2 + x_1^2) + \sin(x_2^2 + x_3^2) \right). \label{last} 
 \end{equation}  

$ x _ i $ in all cases is $ 2 _1 {\bf x}_{ i }^m -1 $. Here, $ _n {\bf x} ^m_{ 2 i } = 0.5 ( \langle \tilde{ \Psi }  ( _1 {\bf x} ^m ) | Z_{ 2 i } | \tilde{ \Psi } ( _1 {\bf x} ^m ) \rangle + 1)$ and $ _n {\bf x} ^m_{ 2 i + 1 } = 0.5 ( \langle \tilde{ \Psi }  ( _1 {\bf x} ^m ) | Y_{ 2 i + 1 } | \tilde{ \Psi } ( _1 {\bf x} ^m ) \rangle + 1)$ which the range is normalized in $ \{ 0, 1 \} $ for the state calculated by n-th layer $ | \tilde{ \Psi } ( _1 {\bf x} ^m ) \rangle $, with $N_d^n = 4$ and $\dim(_n {\bf x} ^m ) = 4$ for all layers and calculations, and the Hamiltonian is $Z_0 Z_1$.   
The initial state for the fitting problem packs two input components onto each qubit,
\begin{equation}
\mid \Psi_{ini} (_1 {\bf x} ^m) \rangle = \prod _{j = 0} ^{ N_d / 2 - 1 } Ry^j (\arccos( _1 {\bf x}_{ 2 j }^m )) \, Rx^j (\arccos( _1 {\bf x}_{ 2 j + 1 }^m )) \mid 0 \rangle ^{\otimes N _q},
\end{equation}
so that with $ N_d = 4 $ the encoding occupies qubits $ 0 $ and $ 1 $ only. Qubit $ 2 $ is initialised in $ \mid 0 \rangle $ and carries no input; it participates in the ansatz as a control and is not read out by the Hamiltonian $ Z_0 Z_1 $. Note that the product runs to $ N_d / 2 - 1 $ and not to $ N_q - 1 $, since the latter would require $ 2 N_q = 6 $ input components.
    
In advance, we show the result for QNN, VQKAN, Adaptive VQKAN\cite{2025arXiv250321336W}, respectively. 
    
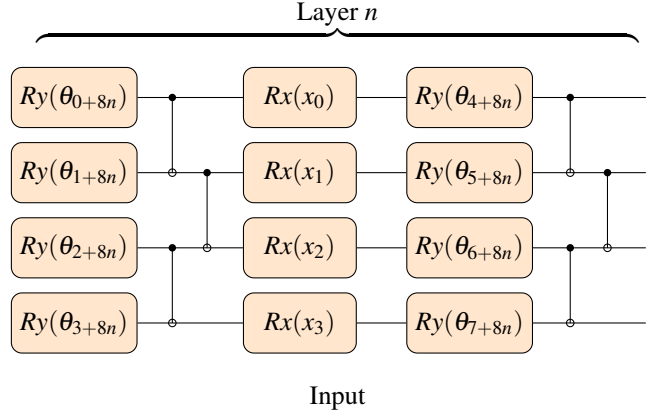
\begin{figure*}[h!] 
 \centering     
 \begin{tikzpicture}[  
 gate/.style={draw, rounded corners, minimum width=1.5cm,  
 minimum height=0.8cm, fill=orange!20}, 
 dot/.style={circle, fill, inner sep=1pt}, 
 control/.style={circle, draw, inner sep=1pt}, 
] 
 
\node at (3.5, 5) {$\overbrace{\hspace{8cm}}^{\scalebox{1}{\makebox[8cm]{\text{Layer} $n$}}}$};

\node[gate] (ry0) at (0, 4) {$Ry(\theta_{0+8n})$};
\node[gate] (ry1) at (0, 3) {$Ry(\theta_{1+8n})$};
\node[gate] (ry2) at (0, 2) {$Ry(\theta_{2+8n})$};
\node[gate] (ry3) at (0, 1) {$Ry(\theta_{3+8n})$};

\node[gate] (rx0) at (3, 4) {$Rx(x_0)$};
\node[gate] (rx1) at (3, 3) {$Rx(x_1)$};
\node[gate] (rx2) at (3, 2) {$Rx(x_2)$};
\node[gate] (rx3) at (3, 1) {$Rx(x_3)$}; 

\node[gate] (ry4) at ($(rx0.east) + (1.5,0)$) {$Ry(\theta_{4+8n})$};
\node[gate] (ry5) at ($(rx1.east) + (1.5,0)$) {$Ry(\theta_{5+8n})$};
\node[gate] (ry6) at ($(rx2.east) + (1.5,0)$) {$Ry(\theta_{6+8n})$};
\node[gate] (ry7) at ($(rx3.east) + (1.5,0)$) {$Ry(\theta_{7+8n})$};
 
\draw (ry0.east) -- (rx0.west);
\draw (ry1.east) -- (rx1.west);
\draw (ry2.east) -- (rx2.west);
\draw (ry3.east) -- (rx3.west);

\draw (rx0.east) -- (ry4.west);
\draw (rx1.east) -- (ry5.west);
\draw (rx2.east) -- (ry6.west);
\draw (rx3.east) -- (ry7.west);

\draw (ry4.east) -- ++(1.5,0) coordinate (ext4);
\draw (ry5.east) -- ++(1.5,0) coordinate (ext5);
\draw (ry6.east) -- ++(1.5,0) coordinate (ext6);
\draw (ry7.east) -- ++(1.5,0) coordinate (ext7);

\node[dot] at ($ (ry0.east)!0.33!(rx0.west) $) {};
\node[control] at ($ (ry1.east)!0.33!(rx1.west) $) {};
\draw ($ (ry0.east)!0.33!(rx0.west) $) -- ($ (ry1.east)!0.33!(rx1.west) $);

\node[dot] at ($ (ry1.east)!0.66!(rx1.west) $) {};
\node[control] at ($ (ry2.east)!0.66!(rx2.west) $) {};
\draw ($ (ry1.east)!0.66!(rx1.west) $) -- ($ (ry2.east)!0.66!(rx2.west) $);

\node[dot] at ($ (ry2.east)!0.33!(rx2.west) $) {};
\node[control] at ($ (ry3.east)!0.33!(rx3.west) $) {};
\draw ($ (ry2.east)!0.33!(rx2.west) $) -- ($ (ry3.east)!0.33!(rx3.west) $);

\node[dot] at ($ (ry4.east)!0.33!(ext4) $) {};
\node[control] at ($ (ry5.east)!0.33!(ext5) $) {};
\draw ($ (ry4.east)!0.33!(ext4) $) -- ($ (ry5.east)!0.33!(ext5) $);

\node[dot] at ($ (ry5.east)!0.66!(ext5) $) {};
\node[control] at ($ (ry6.east)!0.66!(ext6) $) {}; 
\draw ($ (ry5.east)!0.66!(ext5) $) -- ($ (ry6.east)!0.66!(ext6) $);

\node[dot] at ($ (ry6.east)!0.33!(ext6) $) {}; 
\node[control] at ($ (ry7.east)!0.33!(ext7) $) {};
\draw ($ (ry6.east)!0.33!(ext6) $) -- ($ (ry7.east)!0.33!(ext7) $);

\node at (3.5, 0.0) {Input};

\end{tikzpicture}
 \caption{Ansatz structure of the Quantum Neural Network (QNN) with parameterized rotation gates controlled by trainable parameters \(\theta_j \). Inputs \(x_i \) encode classical data into the quantum circuit, enabling learning through optimization of \(\theta_j \) to minimize the cost function.}
 \label{qnn}
\end{figure*}

The QNN ansatz consists of three layers, as shown in Fig.  \ref{qnn}, with a total of 24 parameters initialized randomly. The initial state is $ \mid 0 \rangle ^{\otimes N_q} $.   
      
\begin{figure*}  
\centering  

\includegraphics[scale= 0.25 ]{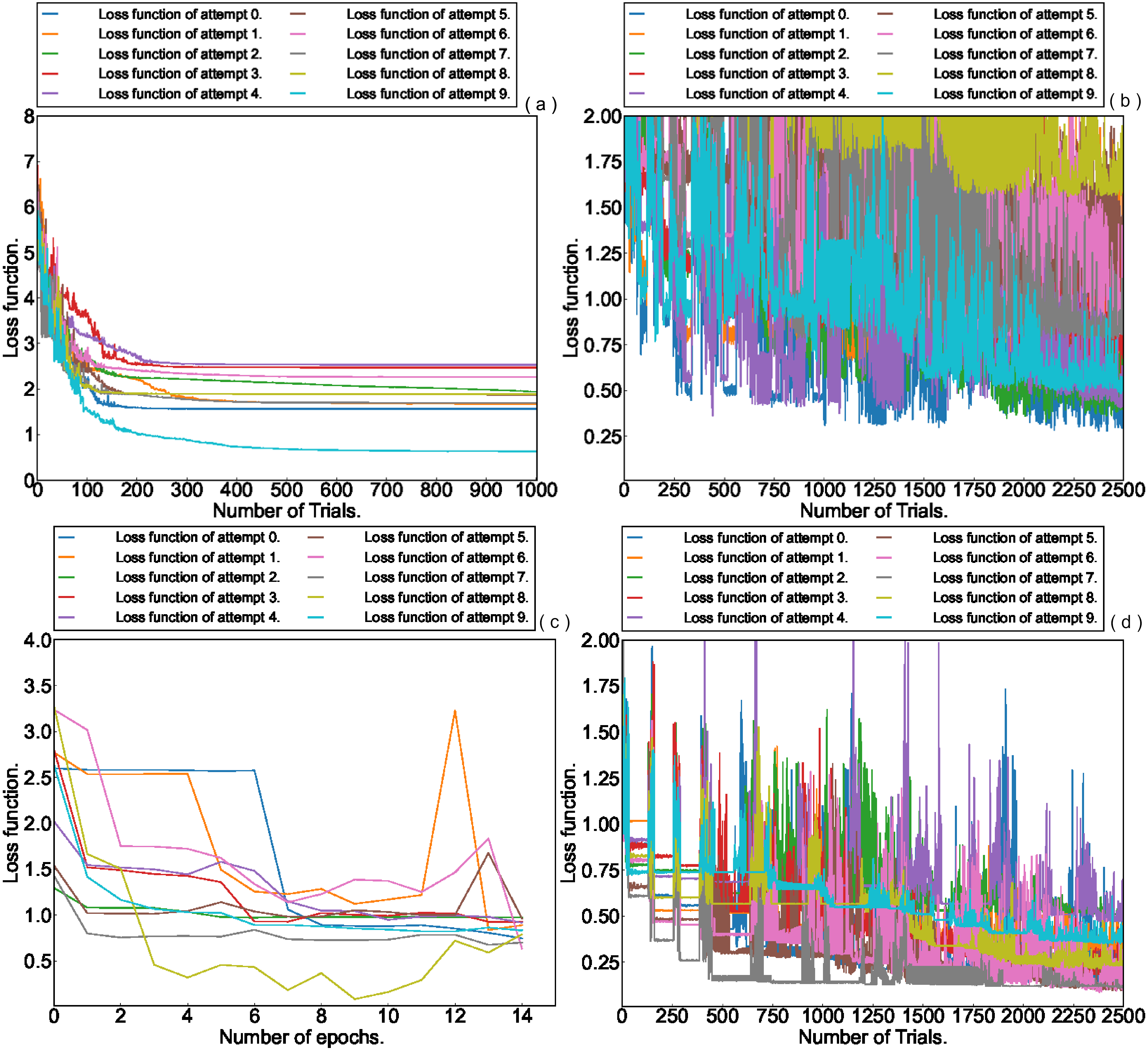}

\caption{Number of trials vs. loss functions for optimization of the fitting problem by ( a ) QNN, ( b ) VQKAN, ( c ) Adaptive VQKAN, and ( d ) Enhanced VQKAN, respectively.} \label{cninn}              
\end{figure*}

\begin{figure*}  
\centering  
   
\includegraphics[scale = 0.5 ]{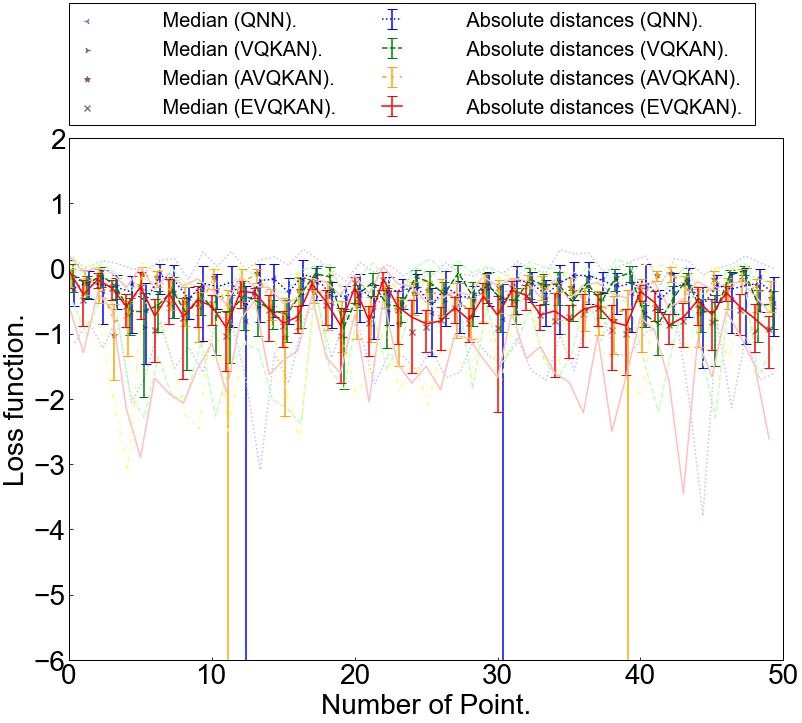}

\caption{(Right) Number of test points vs. average and median of loss functions (absolute distances) in log10 scale of test points on ( blue line ) QNN, ( green line ) VQKAN, ( orange line ) Adaptive VQKAN, and ( red line ) Enhanced VQKAN optimization, respectively. The line of QNN, VQKAN, and Adaptive VQKAN, are moved right 0.375, 0.25, and 0.125, respectively.   } \label{cnin}  

\end{figure*}

Fig. \ref{cninn} ( a ) shows the convergence of the loss function over the number of trials for 10 attempts, while Fig. \ref{cnin} ( blue line ) presents the fitting results on the test points of QNN. 
 
The loss functions of half of ten attempts reached below 0.5. 
However, the average sum of the difference between the absolute value of aimed and calculated value of points ( absolute distances ) is over 25 as also shown in Table.\ref{ fit }. 
According to the values of the loss function, QNN is not good at fitting the equation using the same encoding as EVQKAN and is trapped by overfitting because the average of the absolute distances on each point is nearly 1 even though their minimums are small.

Fig. \ref{cninn} ( b ) and ( c ) shows the convergence of the loss function over the number of trials for 10 attempts, while Fig. \ref{cnin} ( green line ) and ( orange line ) presents the fitting results on the test points of VQKAN and Adaptive VQKAN, respectively.   

The ansatz of VQKAN is 3 layer canonical ansatz and the initial absatz of Adaptive VQKAN is $ X _ 0 $, respectively.
The initial parameters are all zero and $ N _ g = 8 $ the same as EVQKAN, and the number of epochs of Adaptive VQKAN is 15 which the number of trials of optimizer for each epoch is 1000, respectively. 
Absolute distances are entirely smaller than those of QNN and a little larger than those of EVQKAN even though the loss functions are larger than that of QNN.

Next, we show the  result of fitting by EVQKAN on Figs. \ref{cninn} ( d ) and \ref{cninn} ( red line ).            
       
Fig. \ref{cninn} ( d ) shows the convergence of the loss function over the number of trials for 10 attempts, while Fig. \ref{cnin} ( red line ) presents the fitting results on the test points.    
The values loss functions on seven of ten attempts are below 0.5 even though they had not converged yet.        
The average of the sum of the absolute distances is nearly 15 which is about 10 smaller than that of QNN.   
The result of fitting on test points shows a little overfitting because the absolute distances on some test points are below 0.01, and some are above 0.1, even though the averages on test points are entirely smaller than those of QNN.         
The EVQKAN is supposed to be good at predicting learned points in optimization.     
   
\begin{table}[h]
 \caption{
Sum of absolute distances over the 50 test points on the fitting problem, for QNN, VQKAN, Adaptive VQKAN, EVQKAN, classical KAN and LSTM, with $ N_l = 3 $ for all methods, over 10 attempts using COBYLA. Uncertainties are sample standard deviations and brackets give the 95\% confidence interval of the mean. Raw per-attempt values for the four quantum methods are provided with the code.
}\label{ fit }
\centering
\begin{tabular}{c|c|c|c|c|c} \hline \hline
 Method & Mean $ \pm $ SD & 95\% CI & Med. & Min. & Max. \\\hline
QNN&25.965 $ \pm $ 7.027&[20.94, 30.99]&25.688&14.535&35.562 \\\hline
VQKAN&22.609 $ \pm $ 1.668&[21.42, 23.80]&22.957&19.876&24.682 \\\hline
AVQKAN&22.381 $ \pm $ 2.972&[20.25, 24.51]&21.934&16.904&28.294 \\\hline
EVQKAN&{\bf 15.088} $ \pm $ 1.739&[13.84, 16.33]&15.119&13.121&18.640 \\\hline
KAN&13.497&---&13.549&9.588&18.891 \\\hline
LSTM&18.455&---&17.128&16.302&24.840 \\\hline
\end{tabular}
\end{table}

\begin{table}[h]
 \caption{
Significance of the differences in Table \ref{ fit }. Two-sided Mann-Whitney $ U $ test over the 10 attempts of each method, with Cliff's $ \delta $ as the effect size; negative $ \delta $ favours EVQKAN and $ \delta = -1 $ means every EVQKAN attempt beat every attempt of the comparison method.
}\label{ fitstat }
\centering
\begin{tabular}{c|c|c|c} \hline \hline
 Contrast & $ \Delta $ mean & $ p $ & Cliff's $ \delta $ \\\hline
EVQKAN vs. QNN&$ -10.877 $&0.0013&$ -0.860 $ \\\hline
EVQKAN vs. VQKAN&$ -7.521 $&0.00018&$ -1.000 $ \\\hline
EVQKAN vs. AVQKAN&$ -7.293 $&0.00025&$ -0.980 $ \\\hline
\end{tabular}
\end{table}

 Table.\ref{ fit } shows the detail of the sum of absolute distances for QNN, VQKAN, EVQKAN, classical KAN, and Long short term memory ( LSTM ) with $ 2 \times 60 $ parameters and Adam as an optimizer for 25 trials.

EVQKAN attains a mean of $ 15.088 \pm 1.739 $, and Table \ref{ fitstat } shows that its advantage over each quantum baseline is statistically significant: $ p = 0.0013 $ against QNN, $ p = 0.00018 $ against VQKAN and $ p = 0.00025 $ against Adaptive VQKAN. The effect sizes are large in every case, and against VQKAN the separation is complete --- $ \delta = -1.000 $, meaning every one of the ten EVQKAN attempts produced a smaller error than every one of the ten VQKAN attempts. EVQKAN is also more accurate than LSTM.

The comparison with QNN deserves a remark on variance rather than on means alone. QNN has by far the widest spread of the four ( $ \mathrm{SD} = 7.027 $ against $ 1.739 $ for EVQKAN ), with three of ten attempts exceeding 30 and a best attempt of 14.535 that is competitive with EVQKAN's mean. EVQKAN is thus not only more accurate on average but markedly more reliable from run to run, which matters in a setting where one cannot afford many restarts.

Classical KAN, at 13.497, remains more accurate than EVQKAN. We state this plainly: on this task the quantum model does not match the classical model it emulates, and the result reported here is a separation among quantum methods.

The fitting task was optimised on only $ N = 10 $ sampled points against 96 trainable parameters. Section \ref{4} takes up the question of whether this is what limits the accuracy.

\subsection{ Classification problem }

We next apply EVQKAN to a two-dimensional classification task. A point $ ( x_0, y ) $ carries the label $ +1 $ if it lies above the curve $ f $ and $ -1 $ if it lies below, where
\begin{equation}
f(x) = \exp(d_0 x_0 + d_1) + d_2 \sqrt{1-d_3 x_0^2} + \cos(d_4 x_0 + d_5) + \sin(d_6 x_0 + d_7) \label{last c}
\end{equation}
and the $ d_k $ are random coefficients in $ [0,1] $ drawn once per seed. For every sampled $ x_0 $ the generator emits one point above the curve and one below, so the labels are exactly balanced by construction.

\subsubsection*{ Protocol }

The results in this subsection supersede those of an earlier version of this work \cite{Wakaura2025_EVQKAN_v3}, which are withdrawn; Section \ref{4} explains why. Two changes to the protocol matter.

First, every method is now run through a single shared harness. Previously each method had its own driver, and the drivers did not agree on which columns of the sample array were features: the EVQKAN driver encoded all three columns, so the target label reached the circuit as an input, while the VQKAN driver excluded it. Here all arms share the same three-qubit working register, the same encoding of the two features and only the two features, the same observable $ Z_0 $, the same loss $ | y_m - \langle Z_0 \rangle | $ with the same weighting $ a_m $, the same data and the same optimiser. Only the ansatz differs.

Second, we report standard classification metrics --- accuracy, $ F_1 $ and the area under the ROC curve --- alongside the summed distance $ \sum_m | y_m - \langle Z_0 \rangle | $ used previously. On balanced labels that distance has fixed reference points: $ 0 $ for a perfect predictor, $ 50 $ for the constant $ \langle Z_0 \rangle = 0 $ predictor and $ 100 $ for a perfectly wrong one. It is not, however, a measure of classification quality, because it rewards confident output as much as correct output. Table \ref{ classnew } contains a direct demonstration: logistic regression classifies more accurately than every quantum method ( $ 0.790 $ ) yet scores a \emph{worse} summed distance ( $ 43.84 $ ) than VQKAN ( $ 40.08 $ ), simply because its predicted probabilities sit near $ 0.5 $. We therefore treat accuracy and AUC as the primary metrics.

Because COBYLA requires $ n+1 $ evaluations merely to construct its initial simplex, a fixed iteration budget would penalise whichever ansatz carries more parameters. We allocate $ \max ( 400, 8 n ) $ iterations per arm, so QNN receives 400, EVQKAN 768 and VQKAN 1728. Ten independent seeds are run per arm, each trained on 10 points and evaluated on 50 held-out points.

\subsubsection*{ Results }

Table \ref{ classnew } gives the outcome and Table \ref{ classnewstat } the pairwise tests.

All three quantum methods classify above chance: EVQKAN reaches an accuracy of $ 0.620 $ against a chance level of $ 0.5 $ ( one-sample $ t $-test, $ p = 0.0005 $ ). The ordering among them, however, runs against EVQKAN. QNN attains $ 0.754 $ accuracy and $ 0.893 $ AUC, and the gap to EVQKAN is significant on both metrics ( $ p = 0.0014 $ and $ p = 0.0002 $ ). EVQKAN and VQKAN are statistically indistinguishable ( $ p = 0.30 $ and $ p = 0.12 $ ). QNN achieves this with 18 trainable parameters against EVQKAN's 96 and VQKAN's 216.

We tested the most obvious explanation for the discrepancy with the fitting task and it did not hold. The classification variant of the ansatz feeds its four splines from the readout vector $ [ \langle Z_0 \rangle, \langle Z_1 \rangle, \langle Z_0 \rangle, \langle Z_1 \rangle ] $, which contains only two distinct values, whereas the fitting variant uses $ [ \langle Z_0 \rangle, \langle Y_0 \rangle, \langle Z_1 \rangle, \langle Y_1 \rangle ] $ and contains four. Restoring the four-valued readout ( row EVQKAN-ZY ) changes nothing: accuracy $ 0.596 $ against $ 0.620 $, $ p = 0.36 $, and AUC $ 0.674 $ against $ 0.641 $, $ p = 0.79 $. The degenerate readout is not the cause.

Both classical references outperform every quantum method on accuracy, by roughly 15 percentage points over EVQKAN.

The honest summary is that the advantage the tiled ansatz confers on function fitting does not transfer to this classification task. We have not identified the mechanism, and we regard this as the most important open question raised by the present work: an architecture whose benefit is confined to one task family is of limited interest, and establishing whether the boundary is intrinsic or an artefact of the encoding would materially change how EVQKAN should be understood.

\begin{table}[h]
\caption{Leak-free classification comparison. Every arm uses the same encoding, observable, loss, data and optimiser; only the ansatz differs. Ten seeds per arm, 10 training points, 50 held-out test points, uncertainties are sample standard deviations. LogReg and MLP are classical references trained on the same features and split. The summed distance column is retained for comparability with earlier work; see the text for why it should not be read as a measure of classification quality.}\label{ classnew }
\centering
\begin{tabular}{c|c|c|c|c|c|c} \hline \hline
Method & Par. & Iter. & Accuracy & $ F_1 $ & AUC & $ \sum | y - \langle Z_0 \rangle | $ \\\hline
QNN&18&400&{\bf 0.754} $ \pm $ 0.07&0.737 $ \pm $ 0.07&{\bf 0.893} $ \pm $ 0.03&33.45 $ \pm $ 3.43 \\\hline
EVQKAN&96&768&0.620 $ \pm $ 0.07&0.503 $ \pm $ 0.11&0.641 $ \pm $ 0.11&42.89 $ \pm $ 4.15 \\\hline
EVQKAN-ZY&96&768&0.596 $ \pm $ 0.05&0.454 $ \pm $ 0.14&0.674 $ \pm $ 0.06&43.99 $ \pm $ 2.91 \\\hline
VQKAN&216&1728&0.658 $ \pm $ 0.05&0.710 $ \pm $ 0.04&0.735 $ \pm $ 0.09&40.08 $ \pm $ 3.36 \\\hline
LogReg&---&---&0.790 $ \pm $ 0.05&0.777 $ \pm $ 0.05&0.866 $ \pm $ 0.05&43.84 $ \pm $ 1.01 \\\hline
MLP&---&---&0.802 $ \pm $ 0.08&0.805 $ \pm $ 0.07&0.886 $ \pm $ 0.06&20.75 $ \pm $ 7.80 \\\hline
chance&---&---&0.500&---&0.500&50.00 \\\hline
\end{tabular}
\end{table}

\begin{table}[h]
\caption{Pairwise comparison of the arms in Table \ref{ classnew }, two-sided Mann-Whitney $ U $ over the ten seeds. Negative differences favour the first-named method.}\label{ classnewstat }
\centering
\begin{tabular}{c|c|c|c|c} \hline \hline
Contrast & $ \Delta $ acc. & $ p $ (acc.) & $ \Delta $ AUC & $ p $ (AUC) \\\hline
EVQKAN vs. QNN&$ -0.134 $&{\bf 0.0014}&$ -0.252 $&{\bf 0.0002}\\\hline
EVQKAN vs. VQKAN&$ -0.038 $&0.3042&$ -0.093 $&0.1211\\\hline
EVQKAN vs. EVQKAN-ZY&$ +0.024 $&0.3605&$ -0.032 $&0.7913\\\hline
QNN vs. VQKAN&$ +0.096 $&{\bf 0.0062}&$ +0.159 $&{\bf 0.0011}\\\hline
\end{tabular}
\end{table}

\clearpage

\newpage

\section{ Discussion } \label{4}  
  
\begin{figure*}   
\centering 
\includegraphics[ scale=  0.25 ]{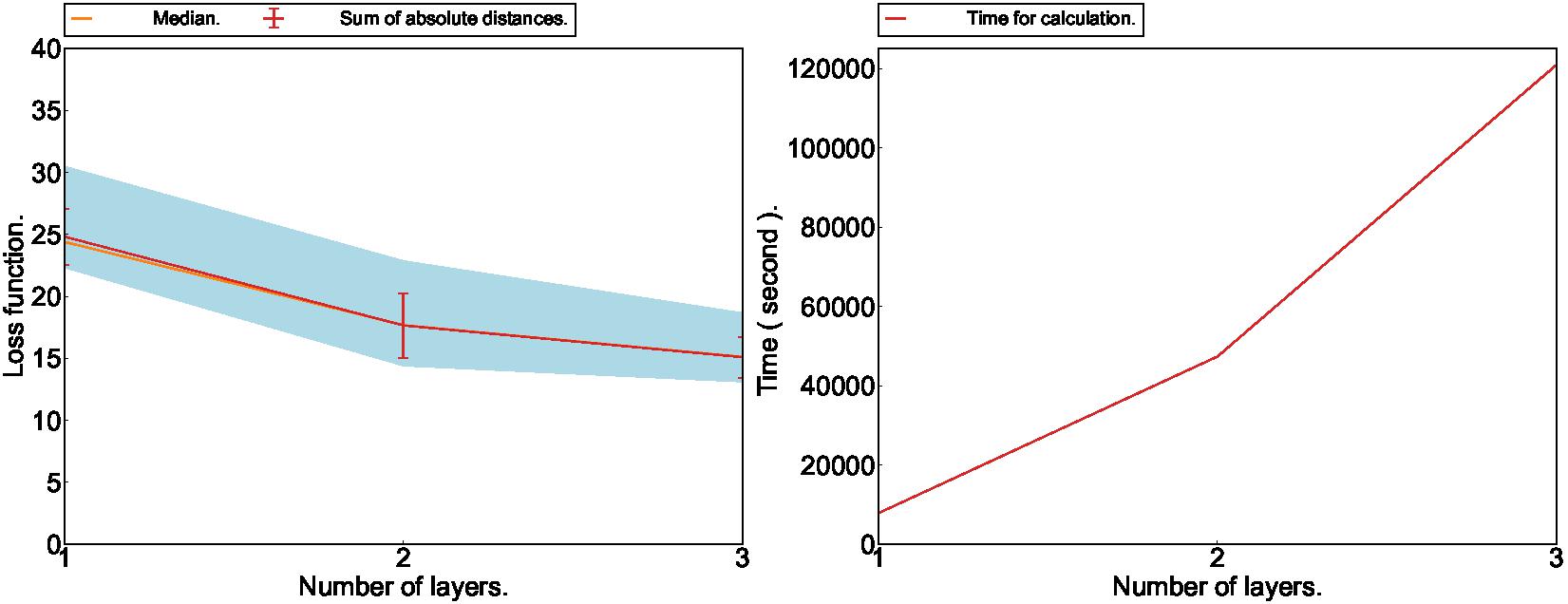}  
 
\caption{(Left) Number of layers vs. sum of absolute distances using EVQKAN on fitting problem with $ N_q = 3 $ and COBYLA.(Right) Number of layers vs. elapsed time. }\label{ fit l }  
        
\end{figure*}

In this section, we discuss the key findings in this work, focusing on the accuracy and time required to calculate EVQKAN for the fitting problem, and on what the classification result implies.  
Firstly, we discuss the accuracy and time of fitting problems, including the reason and the way to improve them. 
Fig.\ref{ fit l } (Left) shows the average of the sum of absolute distances of 50 test points; those are the result of prediction after optimization of EVQKAN on the fitting problem for the number of layers, while (Right) shows the calculation time.  
The average is a little smaller than that of QNN when the number of layers is 1 and declines drastically as the number of layers increases.
The average may be nearly zero when the number of layers is above 6. 
However, the calculation time increases nearly exponentially before saturating. The reason is the gate cost of the sum-operator construction, which we now quantify exactly, since this is the principal practical limitation of the method.

Counting the operations the ansatz emits for $ N_q = 3 $, a single EVQKAN layer contains 339 operations, of which 228 are Toffoli gates, together with 74 $ X $, 33 $ Ry $, 2 $ H $ and 2 controlled-$ Ry $ gates. Three stacked layers --- the configuration of Table \ref{ fit } --- emit 1017 operations including 684 Toffoli gates, on 8 qubits including ancillae. Decomposing each Toffoli into six CNOTs and nine one-qubit gates, the three-layer circuit requires 4110 two-qubit gates and 10593 gates in total. The multi-controlled Toffoli helper alone accounts for 7 Toffoli gates per invocation and is called eight times per tile.

This is a large circuit by any current standard, and it must be weighed against the parameter economy that motivates the construction. EVQKAN trains $ 2^{N_q-1} N_g $ parameters per layer, which for the present configuration is 32 per layer and 96 in total; Quantum KAN instead trains a number of parameters proportional to the number of elements of the layer matrix. The saving is therefore in the classical optimisation problem, not in the quantum circuit. A circuit of 4110 two-qubit gates cannot be executed coherently on present-day hardware, and we make no claim that EVQKAN as constructed here is deployable on NISQ devices. Peephole optimisation of the emitted circuit --- cancelling the adjacent Toffoli pairs inside each tile and folding the $ X $-conjugation patterns --- would reduce these figures by a modest factor, but not by the order of magnitude that hardware execution would require.

\begin{figure*}  
\centering 

\includegraphics[scale=  0.25 ]{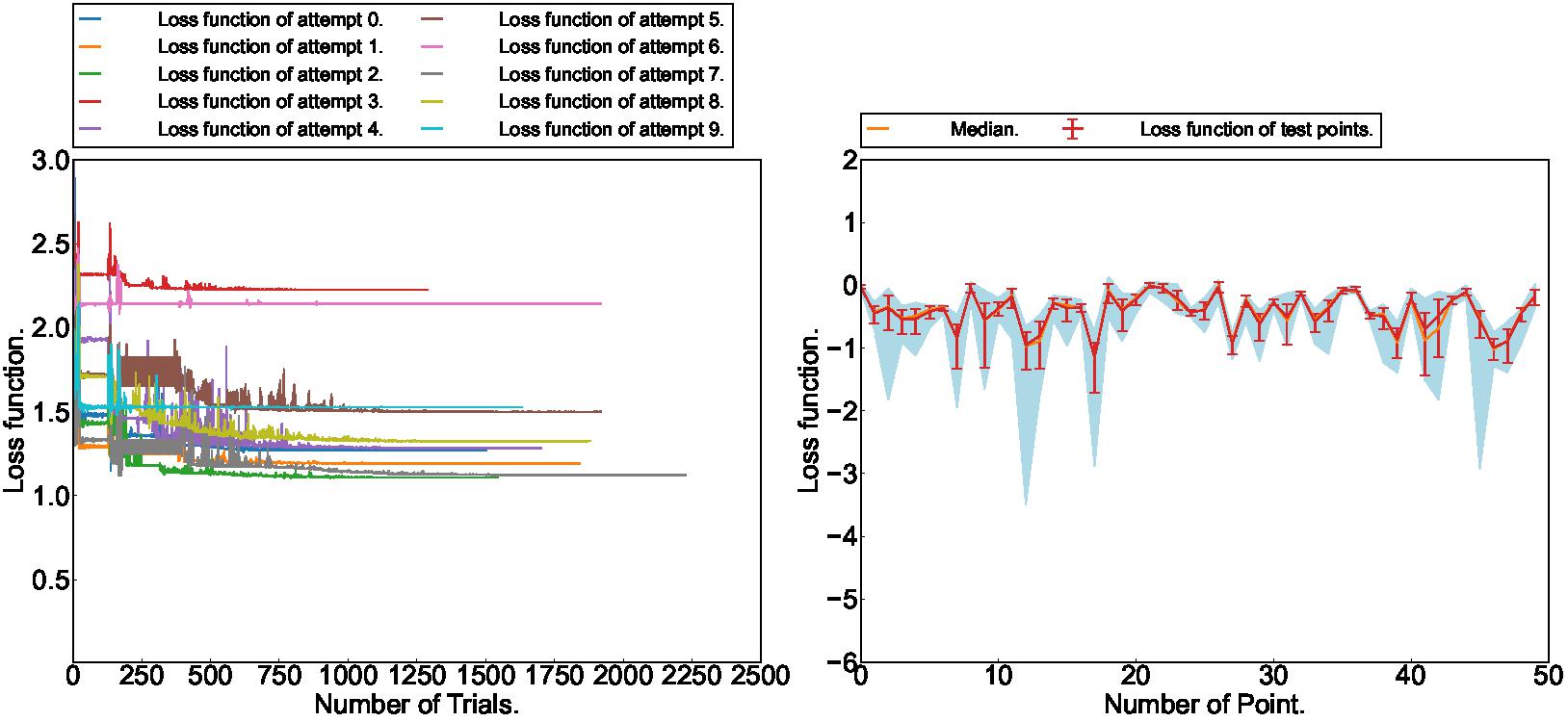}

\caption{
(Left) Number of trials vs. loss functions for optimization attempts on the fitting problem by EVQKAN in case the ansatz is transposed and the number of layers is 1. (Right) Number of test points vs. average and median of loss functions (absolute distances) in log10 scale of test points on EVQKAN optimization.} \label{ e n fit t }    
   
\end{figure*}

Reducing this cost is the central open problem for the method. Block encoding and qubitization simplify the circuit for computing matrix functions on quantum computers and are the natural tools; tensor product decomposition \cite{2022arXiv220702851S} may also accelerate the calculation and reduce the qubit count. Establishing how the gate count of EVQKAN scales with $ N_q $, which we have not done here, is a prerequisite for judging whether the parameter economy survives at larger layer dimensions.
Our ansatz includes one input vector element per row. Hence, the state vectors of EVQKAN include only one input vector element per element. 
The transposed ansatz includes all their elements; hence, the accuracy of EVQKAN can improve.  
Fig. \ref{ e n fit t } (Left) shows the convergence of the loss function over the number of trials for 10 attempts in case the ansatz is transposed and the number of layers is only 1, while Fig. \ref{ e n fit t } (Right) presents the fitting results on the test points.   
  
Though the loss functions are larger than that of ordinary ansatz, the absolute distances of test points have smaller values entirely than those of ordinary ansatz, even if the number of layers is only 1.  
    
Furthermore, the average of the sum of the absolute distances on fitting of is smaller than that in case the ansatz is ordinary and the number of layers is 1, and QNN in case the number of layers is 3 as shown in Tables.\ref{ fit }, \ref{ ab di l } and Fig. \ref{ fit l } (Left). In addition, they are as small as those of VQKAN and Adaptive VQKAN. 
  
EVQKAN of transposed ansatz has more prediction accuracy than QNN on fitting of logarithmic and radius of sphere which the center is zero point.  
However, it has low prediction accuracy than QNN on exponential and fractional function. 
It is supposed to be because normalized exponential function is almost flat due to too large maximum and the number of layer is too small to optimize accurately.

\begin{table}[h] 
\caption{Averages, medians, minimums, and maximums of the sum of absolute distances of EVQKAN which the number of layers is 1 and ansatz is transposed and QNN which the number of layers is 3 for eq. \ref{last}, exponential function, logarithmic function, fractional function and radius from zero point for 10 attempts using COBYLA. The range of $ x _ i $ is the same as that of eq. \ref{last}. }\label{ ab di l }

\centering            
\begin{tabular}{c|c|c|c|c|c} \hline \hline    
 & & Ave. & Med.& Min. & Max.  \\\hline     
Eq. \ref{last}&EVQKAN &15.088392&15.118825&13.120516&18.640238 \\\hline
&QNN&25.965271&25.688473&14.535198&35.561974 \\\hline 
$ exp (  ( x _ 1 -  x _ 2 ) ^ 2  / 2 x _ 0 )  $&EVQKAN&29.217728&28.199258&25.958048&34.785128 \\\hline    
 &QNN&26.846101&25.881893&17.682263&40.463479 \\\hline   
$ log ( x _ 0 / x _ 1 )   $&EVQKAN&1.957241&2.005758&1.255589&2.670427 \\\hline 
&QNN&11.266087&11.300752&8.653199&14.404135 \\\hline         
$ 1 / ( 1 + x _ 0 x _ 1 ) $  &EVQKAN& 27.452014&27.740392&24.718224&29.842958 \\\hline   
 &QNN&20.399586&20.862449&12.623779&24.732792 \\\hline         
$ \sqrt { x _ 0 ^ 2 + x _ 1 ^ 2 + x _ 2 ^ 2 }     $&EVQKAN& 12.407651&12.595418&10.90763&14.011901 \\\hline    
 &QNN&18.851256&18.31309&14.915787&28.090903 \\\hline  
\end{tabular}           
\end{table}

\subsection{ What limits the accuracy: an under-determined training set }

The experiment in Section \ref{3} was optimised on $ N = 10 $ sampled points against 96 trainable parameters, that is 9.6 parameters per sample. The overfitting we observed is what one would expect from a model fitted in that regime, and the question is whether it is a property of the ansatz or of the experimental protocol. To settle this we swept the training-set size over $ N \in \{ 10, 25, 50, 100 \} $, scoring each run on a test set of 50 points held fixed within each seed so that the numbers are comparable across $ N $, with four independent seeds per value and all other settings unchanged.

Table \ref{ samplecx } shows the outcome. The generalisation gap --- the difference between test and training mean absolute error --- falls monotonically from 0.243 at $ N = 10 $ to 0.101 at $ N = 100 $, a reduction of $ 58\% $, and the trend against $ \log N $ is strong ( Spearman $ \rho = -0.849 $, $ p = 3 \times 10^{-5} $ over the 16 runs ). The overfitting is therefore driven by the size of the training set and not by the ansatz.

The consequence for test accuracy is more modest, and we are careful not to overstate it. The test mean absolute error falls from 0.335 to 0.305, an improvement of $ 9\% $, improving in three of the four seeds; the trend is marginal ( $ \rho = -0.473 $, $ p = 0.064 $, with a regression slope of $ -0.0080 $ per doubling of $ N $ at $ p = 0.044 $ ). Even at $ N = 100 $, where the model has one training sample per parameter, EVQKAN remains $ 13\% $ less accurate than classical KAN. One caveat applies: the optimiser budget was held fixed at 400 COBYLA iterations across all $ N $, and the training error rises with $ N $ ( from 0.092 to 0.204 ), which is consistent with the larger runs not having converged. The sweep may therefore understate the benefit of additional data.

The practical reading is that enlarging the training set is a real but bounded lever. It reliably closes the train-test gap, and it should be the first thing changed in any future EVQKAN experiment, but on this task it does not by itself bring the quantum model level with its classical counterpart.

\begin{table}[h]
\caption{Training-set size sweep for EVQKAN with $ N_q = 3 $, $ N_l = 3 $ and 96 trainable parameters, over four seeds per row, scored on 50 held-out test points. MAE is the mean absolute error per test point and the gap is the difference between test and training MAE. The last column rescales the test MAE to the sum over 50 points used in Table \ref{ fit }.}\label{ samplecx }
\centering
\begin{tabular}{c|c|c|c|c|c} \hline \hline
$ N $ & par./sample & train MAE & test MAE & gap & test sum \\\hline
10&9.6&0.092 $ \pm $ 0.042&0.335 $ \pm $ 0.019&0.243&16.745 \\\hline
25&3.8&0.161 $ \pm $ 0.031&0.328 $ \pm $ 0.011&0.168&16.415 \\\hline
50&1.9&0.183 $ \pm $ 0.033&0.328 $ \pm $ 0.017&0.145&16.380 \\\hline
100&1.0&0.204 $ \pm $ 0.020&{\bf 0.305} $ \pm $ 0.016&{\bf 0.101}&15.261 \\\hline
KAN&---&---&0.270&---&13.497 \\\hline
\end{tabular}
\end{table}

\subsection{ Limitations }

Three limitations bound the scope of these results and we state them together rather than leaving them implicit.

All calculations are noiseless state-vector simulations with an infinite number of shots. We have not simulated depolarising or dephasing noise, nor finite sampling, and we therefore make no claim about the behaviour of EVQKAN under either. Combined with the gate counts reported above, this means the present work says nothing about NISQ-era performance.

All experiments use $ N_q = 3 $, a single target function per task, and 10 training points. How the accuracy and the gate count of EVQKAN scale with $ N_q $ is untested, and it is the scaling rather than the $ N_q = 3 $ point that would determine whether the construction is useful.

We do not know why the tiled ansatz helps on fitting and hurts on classification. We ruled out the degenerate readout of the classification variant by restoring the four-valued readout, which changed nothing ( $ p = 0.36 $ on accuracy ). The parameter count and the optimiser budget both favour EVQKAN over the QNN that beats it, so neither explains the gap either. Until this is understood, the fitting result should be read as task-specific rather than as a general property of the architecture.

Classical methods outperform every quantum method considered here: classical KAN on the fitting task, and both logistic regression and a small MLP on the classification task. The separation we establish is among quantum models. Demonstrating a regime in which a quantum KAN is preferable to a classical one remains open, and nothing in this paper bears on it.

Finally, a note on scope. An earlier version of this work \cite{Wakaura2025_EVQKAN_v3} also reported a two-dimensional classification experiment. We have since found that the input encoding used in those runs read the number of columns of the sample array directly, and that the classification sample array carries the target label as its third column alongside the two features. The label was therefore encoded into the third qubit as an input feature while also serving as the regression target of the loss. Because the third qubit acts as a control in the tiled rotations, it influences the measured observable: flipping only the encoded label, with the features and parameters held fixed, changes the prediction by $ 0.37 $ on average on a $ [-1,1] $ scale. Retraining with the label removed from the encoding lowers the test AUC from $ 0.79 $ to $ 0.64 $ ( $ p = 0.030 $ over six seeds ). The defect is also asymmetric: the VQKAN and Adaptive VQKAN runs against which EVQKAN was compared excluded the label column correctly, so the comparison did not treat the methods alike. We therefore withdraw those classification results in full and replace them with the leak-free re-run of Section \ref{3}, in which every arm shares one harness and is scored with standard metrics. That re-run reverses the earlier conclusion: EVQKAN is significantly less accurate than QNN rather than comparable to it. The diagnostic scripts that establish all of the above, and the re-run itself, are included in the code release.

 \newpage

\section{Concluding Remarks} \label{7}

In this paper we introduced EVQKAN, a variational ansatz that emulates a $ 2^{N_q} $-dimensional KAN layer by tiling controlled rotations through a sum-operator construction, and that requires only $ 2^{N_q-1} $ trainable spline functions per layer.

On the fitting task EVQKAN is significantly more accurate than QNN, VQKAN and Adaptive VQKAN, with $ p < 0.002 $ in every comparison and effect sizes ranging from large to complete separation over ten independent attempts; it is also markedly more reproducible from run to run than QNN, whose standard deviation is four times larger. On the classification task, re-run under a leak-free protocol that supersedes our earlier report, the ordering reverses: EVQKAN classifies above chance but falls significantly behind QNN on both accuracy and AUC. Classical methods are more accurate than every quantum method we tested on both tasks.

We identified two concrete limits on the method. The accuracy is bounded by overfitting arising from an under-determined training set: enlarging the training set closes the train-test gap by $ 58\% $, though the resulting gain in test accuracy is a more modest $ 9\% $. The circuit cost is dominated by multi-controlled gates, with three layers requiring 4110 two-qubit gates after decomposition, which places the present construction outside the reach of NISQ hardware.

The next objective is therefore to simplify the circuit, including reducing the number of qubits, by block encoding and qubitization, and to establish how the gate count scales with $ N_q $. Benchmarking accuracy and resilience under simulated noise and finite sampling, which we have not attempted here, is equally important before any claim about hardware performance can be made.

 \section*{ Data and code availability }

The implementation of EVQKAN, the scripts that reproduce every table and figure in this paper, and the per-attempt raw values underlying Tables \ref{ fit }, \ref{ fitstat }, \ref{ classnew }, \ref{ classnewstat } and \ref{ samplecx } are openly available at \url{ https://github.com/ } [ repository URL to be inserted on acceptance ] and archived with a DOI at Zenodo. The gate counts quoted in Section \ref{4} are produced by the script \texttt{ resource\_count.py } in that repository, the statistical tests by \texttt{ published\_stats.py }, the label-encoding diagnostics by \texttt{ leakage\_test.py }, and the leak-free classification comparison by \texttt{ classification\_rerun.py }, so that every number reported here can be regenerated from the code.

\section*{ Author Declarations  }

\subsection* { Conflict of Interest  } 

The authors have no conflicts to disclose.  
 
\subsection* {  Author Contributions   }   

Hikaru Wakaura : Conceptualization (lead); Data curation (lead); Formal analysis (lead); Investigation (lead); Methodology (lead); Project administration (equal); Visualization (equal); Writing – original draft (lead); Writing – review$ \& $editing (equal).
Rahmat Mulyawan : Visualization (equal); Writing – review$ \& $editing (equal). 
Andriyan B. Suksmono : Project administration (equal); Writing – review$ \& $editing (equal).

\bibliography{mainwakaura2}

\end{document}